\documentclass[a4paper]{article}
\usepackage[utf8]{inputenc}
\usepackage[T1]{fontenc}
\usepackage{textcomp, mathtools, amssymb, amsthm, wasysym, authblk}
\usepackage[colorlinks=true, linkcolor=red, citecolor=blue, urlcolor=blue]{hyperref}
\usepackage[hmargin=1.5cm, vmargin=1.5cm]{geometry}

\usepackage[backend=biber,style=numeric-comp,sorting=none,maxbibnames=10]{biblatex}
\bibliography{biblio.bib}
\AtEveryBibitem{\clearfield{issn}}
\AtEveryBibitem{\clearfield{isbn}}
\AtEveryBibitem{\clearfield{issue}}
\AtEveryBibitem{\clearfield{doi}}
\AtEveryBibitem{\clearfield{url}}
\AtEveryBibitem{\clearfield{note}}
\AtEveryBibitem{\clearfield{eprint}}
\AtEveryBibitem{\clearfield{day}}
\AtEveryBibitem{\clearfield{month}}

\title{Isotope Ratio Dual-Comb Spectrometer}

\author[1,*]{Alexandre~Parriaux}
\author[1]{Kamal~Hammani}
\author[2,3]{Christophe~Thomazo}
\author[1]{Olivier~Musset}
\author[1,3]{Guy~Millot}

\affil[1]{Laboratoire Interdisciplinaire Carnot de Bourgogne, CNRS UMR 6303, Université Bourgogne Franche-Comté, Dijon, France}
\affil[2]{Biogéosciences, CNRS UMR 6282, Université Bourgogne Franche-Comté, Dijon, France}
\affil[3]{Institut Universitaire de France (IUF), 1 Rue Descartes, Paris, France \linebreak}
\affil[*]{Corresponding author: \href{mailto:alexandre.parriaux@u-bourgogne.fr}{alexandre.parriaux@u-bourgogne.fr}}

\date{}

\begin{document}	
\maketitle

\begin{abstract}
We demonstrate the use of dual-comb spectroscopy for isotope ratio measurements. We show that the analysis spectral range of a free-running near-infrared dual-comb spectrometer can be extended to the mid-infrared by difference frequency generation to target specific spectral regions suitable for such measurements, and especially the relative isotopic ratio $\delta {}^{13}$C. The measurements performed present a very good repeatability over several days with a standard deviation below 2\permil{} for a recording time of a few tens of seconds, and the results are compatible with measurements obtained using an isotope ratio mass spectrometer. Our setup also shows the possibility to target several chemical species without any major modification, which can be used to measure other isotopic ratios. Further improvements could decrease the uncertainties of the measurements, and the spectrometer could thus compete with isotope ratio spectrometers currently available on the market.
\end{abstract}

\section{Introduction}
Isotope ratio measurements (IRMs) is a technique widely used for a large range of applications and hence showing a lot of interest. For instance, strontium ratio $^{87}$Sr/$^{86}$Sr analyses were successfully used to reconstruct the Egtved Girl's life that occurred 3500 years ago~\cite{frei-scirep-2015}. Still related to chronological dating, measurements of the oxygen ratio $^{18}$O/$^{16}$O, carbon ratio $^{13}$C/$^{12}$C and thorium ratio $^{230}$Th/$^{232}$Th ratio have been used for the dating of prehistoric human constructions such as the Bruniquel cave~\cite{jaubert-nature-2016}.
In astronomy, IRMs allow to study the atmosphere and sedimentary rocks of other planets such as Mars~\cite{webster-science-2013,tian-pnas-2021,house-pnas-2022}, or to investigate the composition of meteorites by analyzing magnesium or oxygen isotopes~\cite{ruf-pnas-2017,tartese-pnas-2018}.	
In more daily life related applications, food and beverages can be subjected to IRMs~\cite{camin-revfood-2016}, such as for testing ingredients in beers~\cite{brooks-jafc-2002}, or for detecting the geographical origin of wines~\cite{day-jsfa-1995,magdas-iehs-2012}, mushrooms~\cite{chung-foodchem-2018}, and even cheeses~\cite{camin-foodchem-2004}.
Another example of application where IRMs are performed is the medical domain such as for the detection of Helicobacter Pylori infection, which can be done by measuring the carbon ratio ${}^{13}$C/$^{12}$C out of the exhaled air of a patient~\cite{graham-1987}.

All the examples presented above mostly use a mass spectrometer for IRMs, which is a specific high-performance instrument, but costly and with measurements times that can be of several minutes. Moreover, the samples under study might need to be pre-treated before being analyzed which increases the complexity and can induce more uncertainties~\cite{yang-massspecrev-2009}.
Hence, other techniques for IRMs have been developed, and absorption spectroscopy is one of them when studying gases. This non-destructive technique presents the advantage of being based on optics and lasers which gives an instrument smaller in size and capable to be used for on field measurements~\cite{kerstel-book-2004}. Direct absorption~\cite{becker-appopt-1992}, frequency modulation~\cite{waechter-apb-2007} or cavity ring-down spectroscopy~\cite{wahl-isoenvhealth-2006} are generally used for these measurements.
However, in the field of absorption spectroscopy, new techniques have also been developed in the recent years such as dual-comb spectroscopy (DCS)~\cite{coddington-optica-2016,fortier-commphys-2019,picque-natphot-2019}. DCS shows several characteristics such as a high spectral resolution, no mobile parts in the setup, and the potential to perform real-time analyses.
These features are very interesting for IRMs in gases, and to the best of our knowledge, only a few works have considered the use of DCS for this application~\cite{muraviev-natphot-2018,vodopyanov-wiley-2020}, but no quantitative measurements were performed.

In this paper, we present the use of DCS for IRMs and we demonstrate the feasibility of the technique by measuring the relative carbon isotopic ratio $\delta ^{13}$C of gas samples. First we describe the DCS method and the experimental setup considered here. Then we will see some examples of absorption spectra that can be recorded with our setup to illustrate its features. After that we will present and show how IRMs can be performed with our DCS setup by taking the example of carbon dioxide for $\delta ^{13}$C measurements. In addition, we will show that the frequency tunability of our setup can be used to target other chemical species for IRMs, and especially nitrous oxide for nitrogen isotopes studies.
Finally, we will discuss our results and potential enhancements that could be done in future works to improve the use of the DCS technique for IRMs with a higher accuracy and a lower uncertainty.

\section{Mid-infrared electro-optic dual-comb setup}
Spectroscopic applications can be performed using several techniques, but DCS has been shown to be highly suitable for high resolution spectral measurements and short time acquisitions. This technique, which is based on the study of the interference between two mutually coherent frequency combs with slightly different repetition frequencies, has been intensively studied and refined in the last years~\cite{coddington-optica-2016,fortier-commphys-2019,picque-natphot-2019}.
The Fourier transform of the interference time signal, which is called the interferogram, is also a comb but with a linespacing that is the difference of the repetition frequencies of the combs and with a central frequency that lies in the radio-frequency (RF) domain. When at least one of the optical comb passes through a gas sample, the absorption feature of the gas will be seen in the optical domain but also in the RF domain, which makes it much more easier and straightforward to detect by using only a low bandwidth photodetector and an oscilloscope.
Since the comb linespacing can easily be of the order of the hundred of MHz, using the technique for gas analysis enables high resolution spectroscopic applications.  

\begin{figure}[htb]
\centering
\includegraphics[width=\linewidth]{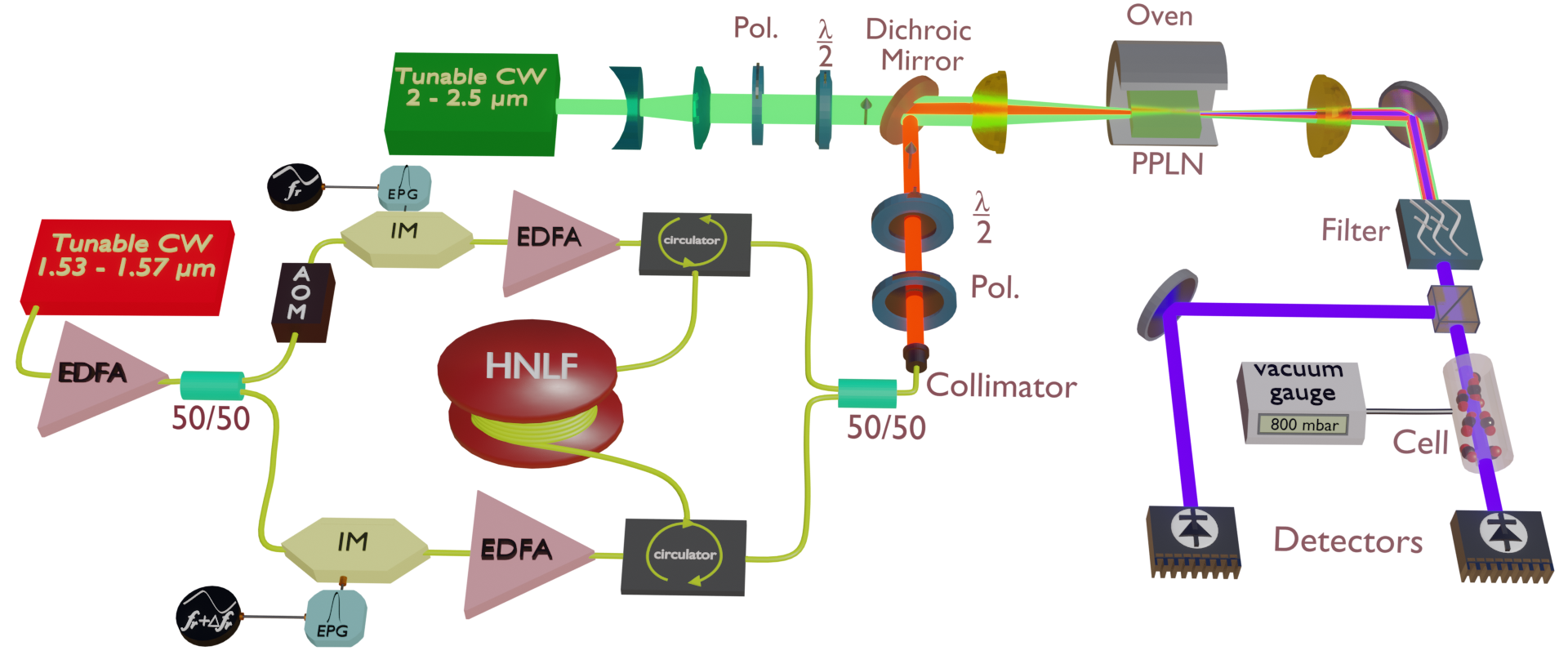}
\caption{Schematic showing the mid-infrared dual-comb spectrometer based on the difference frequency conversion of combs generated by electro-optic modulators around 1.55µm. CW: continuous wave, EPG: electrical pulse generator, EDFA: erbium doped fiber amplifier, AOM: acousto-optic modulator, IM: intensity modulator, HNLF: highly nonlinear fiber.}
\label{fig:setup}
\end{figure}

Lots of experimental setups have been designed for DCS but here, we use a setup based on electro-optic modulators~\cite{parriaux-aop-2020}. This setup, whose architecture is presented in detail in Ref.~\cite{millot-natphot-2016}, shows particular advantages for DCS since there is no need of locking between the combs which simplifies drastically the technique~\cite{millot-natphot-2016,parriaux-aop-2020}.
However, in its basic design~\cite{millot-natphot-2016}, the setup operates around 1.55µm, which is not really suitable for spectroscopic applications. Indeed, most molecules do not show strong absorptions features in this spectral region, thus long absorption cells are required to perform DCS, which can be cumbersome. To bypass this problem, the working spectral region of the setup can be extended to the mid-infrared (MIR) region where molecular absorptions are much stronger than at 1.55µm~\cite{yan-lsa-2017}. In our case, we use difference frequency generation (DFG) in a periodic poled lithium niobate (PPLN) crystal to reach the spectral region between 4.2µm and 4.85µm.
The experimental setup and its spectral extension part to reach the MIR is presented in \autoref{fig:setup}.

The dual-comb spectrometer can be divided into two parts. The first one is the generation of frequency combs at 1.55µm as described in Ref.~\cite{millot-natphot-2016}, and the second part is the frequency conversion of the previous combs in the MIR using DFG. Regarding the part at 1.55µm, we start with a continuous wave (CW) laser at 1.55µm that is amplified up to 150mW. The CW is then split in two arms that are both intensity modulated using electro-optic modulators, which produce 40GHz wide frequency combs with a cardinal sine shape. The intensity modulators are driven by electrical pulse generators which operate by the control of sinusoidal waveform generators that set the repetition frequency of the combs to $ f_r = 250 $MHz and $ f_r + \Delta f_r = 250.025 $MHz. Note that prior to the modulation, one of the arm is frequency shifted by 40MHz using an acousto-optic modulator to prevent aliasing in the later detection.
After generating combs, both arms are amplified up to a peak power of 26W and sent counter-propagatively in a 800m long highly nonlinear fiber of dispersion $D=-91$ps.nm$^{-1}$.km$^{-1}$, linear loss $ \alpha=0.55$dB.km$^{-1}$ and Kerr coefficient $ \gamma=4.6 $W$^{-1}$.km$^{-1}$ at 1.55µm. This fiber was chosen to obtain a spectral broadening as high as possible and a flat-topped shape by dispersive shock waves~\cite{millot-natphot-2016,yan-lsa-2017,parriaux-ol-2017}, giving us spectra that are 3nm wide.

After spectral broadening, both combs are mixed with a 50/50 combiner and sent to free space using a fibered collimator. Using a dichroic mirror, the combs with a peak power of 1.4W are then combined with an idler wave consisting of a wavelength tunable CW laser from 2µm to 2.5µm whose power can be set up to 5W. The beam is then focused on a 10mm long temperature controlled commercially available PPLN crystal (Covesion) for DFG. Note that the polarization of the combs and the idler wave are controlled with polarizers and half-wave plates to be aligned with the dipole moment of the crystal, which is needed to achieve the highest DFG efficiency.
Also note that by modifying appropriately the wavelength of the idler, the temperature and the poling period of the PPLN crystal, the central wavelength of the MIR signal can cover the range 4.2µm to 4.85µm.
Here, we choose to convert both combs in a single crystal which shows a certain ease of doing~\cite{jerez-acsphot-2018}, but note that the combs could be converted individually, which would enable the possibility to perform dispersion spectroscopy~\cite{yan-lsa-2017,luo-pccp-2019}. 
At the output of the crystal, the beam is collimated and filtered to keep only the MIR part. A 50/50 beamsplitter is then used to split the MIR signal in two parts, providing a signal and a reference arm which are both ending with a photovoltaic detector (Vigo) for acquiring the dual-comb interferograms. The signal arm is composed of a pressure monitored cell that can be filled with various gases.

\begin{figure}[htb]
\centering
\includegraphics[]{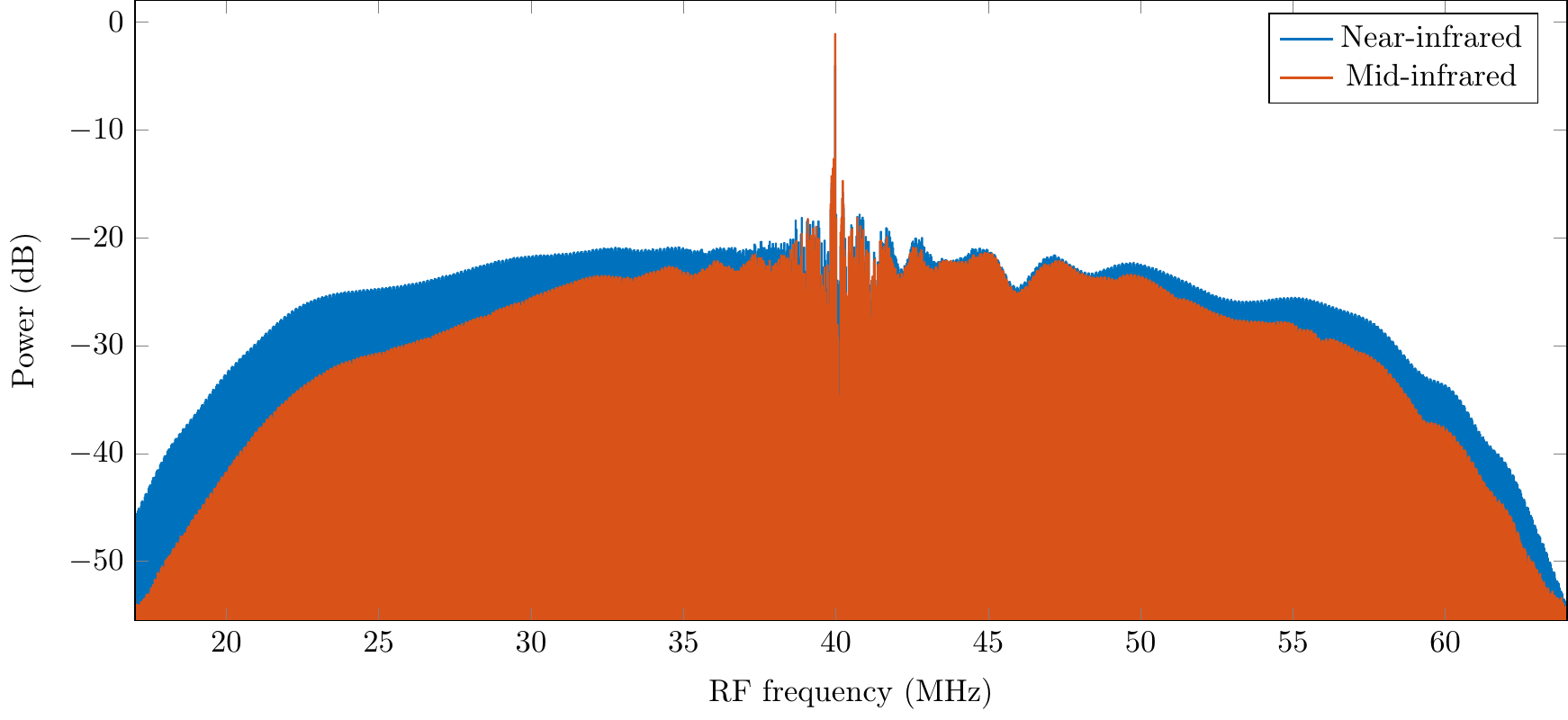}
\caption{Comparison of the radio-frequency combs obtained by the beating of the near-infrared (blue) and mid-infrared (red) combs. The bandwidth difference is due to the finite spectral acceptance of the crystal used for difference frequency generation in the mid-infrared.}
\label{fig:compspecRF}
\end{figure}

We recorded an interferogram of 80ms that is Fourier transformed to reveal the RF comb made of the beating between the MIR combs. This RF comb is compared with the one we can obtained at 1.55µm under similar conditions and both of them are presented in \autoref{fig:compspecRF}. We can observe that the RF comb resulting from the MIR beating is slightly narrower than the RF comb resulting from the near-infrared beating, which is due to the spectral acceptance bandwidth of the PPLN crystal but besides that, both RF combs are highly similar and possess a signal-to-noise ratio of 35dB.

\section{Spectroscopic applications}
\subsection{Absorption spectroscopy}
\begin{figure}[htb]
\centering
\includegraphics[]{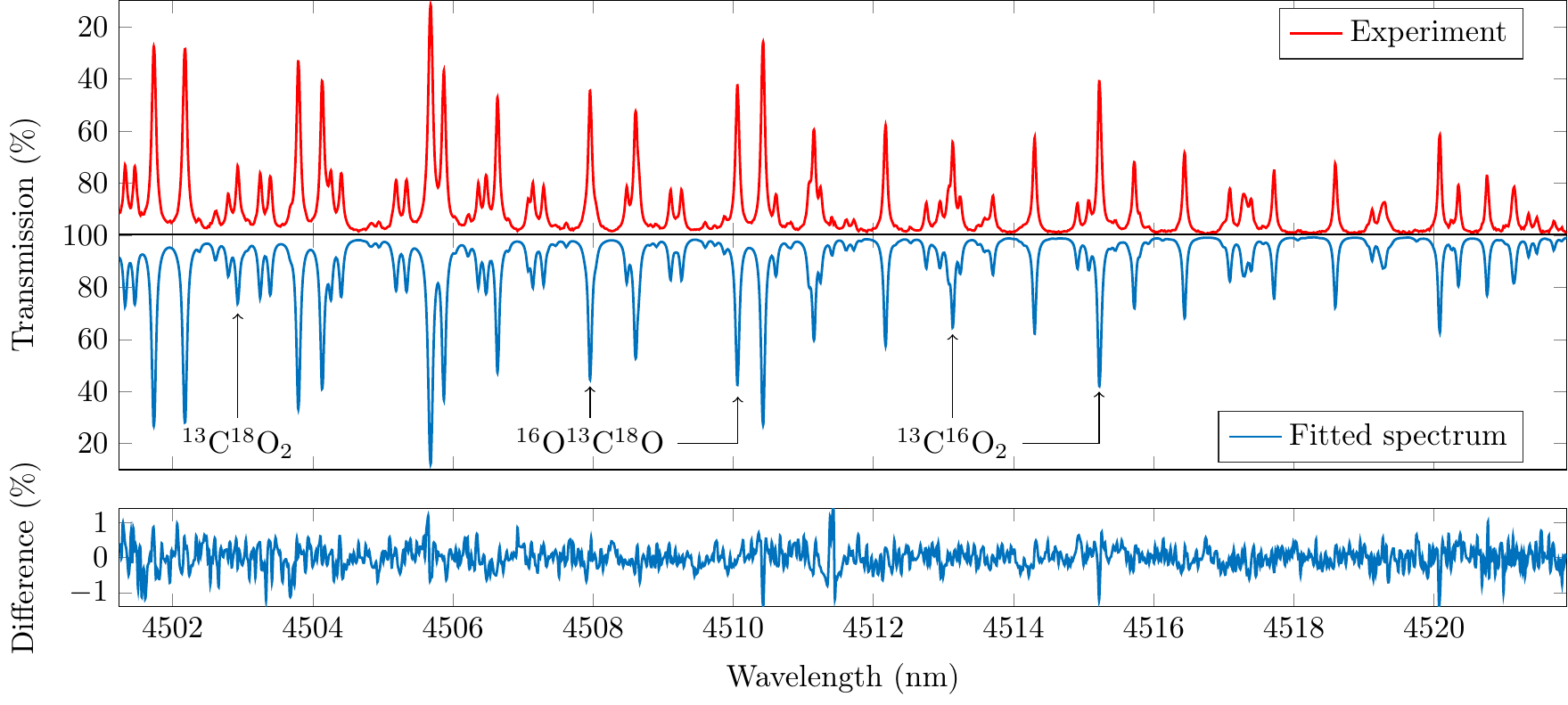}
\caption{(Top) Absorption spectrum of a CO$_2$ gas sample made of 99\% of $^{13}$CO$_2$ recorded around 4511nm at a pressure of 173mbar using a 10s long interferogram. (Middle) Fitted absorption spectrum using line parameters coming from the HITRAN database. (Bottom) Difference between the experimental and fitted spectra.}
\label{fig:full13C}
\end{figure}

Since MIR combs are generated, spectroscopy can be performed. First, we set the wavelength of the near-infrared combs to 1555nm and the idler wave to 2373nm, giving us a MIR signal around 4511nm. We use a 5cm long cell that is filled with a CO$_2$ gas sample made of 99\% of $^{13}$CO$_2$ at a total pressure of 173mbar, and we record at the same time two 10s long interferograms, one for the reference arm and one for the signal arm. Both of them are divided into 80ms sub-interferograms, Fourier transformed and averaged to obtain the reference and signal RF combs.
The envelope of the resulted combs are extracted and compared for baseline correction, which gives us the transmission spectrum of the sample in the cell that is presented in \autoref{fig:full13C}~(top). The frequency scale of the spectrum is converted back in the optical domain using the parameters of our setup and the relation between the RF and optical domains. In the following, the same procedure will be performed to obtain the transmission spectrum of other gas samples under study.

Once the transmission spectrum is obtained, we compared it with a fitted spectrum based on the least-squares regression of a set of Voigt profiles with line parameters given by the HITRAN database~\cite{hitran,hitran2020}. For the regression, the isotopic ratios ${}^{13}r={}^{13}$C/$^{12}$C and ${}^{18}r={}^{18}$O/$^{16}$O are taken as free parameters. Since we are not referencing any of the CW lasers we use, we also take into account a small shift for re-centering the spectrum relatively to the rough frequency value expected by the DFG process. We also consider a baseline correction of the absorption spectrum since the comparison between the reference and signal arm is not perfect.
The obtained fitted spectrum is shown in \autoref{fig:full13C}~(middle) and the difference with the experimental data is presented in \autoref{fig:full13C}~(bottom) showing a very good agreement.

Due to the gas chosen for this experiment and the spectral region investigated, \autoref{fig:full13C} shows several lines coming from different isotopologues where some of them are highlighted. With this simple example, one can imagine that the data obtained from the comparison between the experimental and fitted spectra can be used to extract isotopic ratios such as here, ${}^{13}r$ or ${}^{18}r$, and thus perform IRMs.

\subsection{Isotope ratio measurements}
The basic idea for IRMs using absorption spectroscopy is to target a spectral region where the setup can record the absorption features of several isotopologues. In our case, and compared to the last spectrum we presented, we will be more restrictive since we will focus on particular regions where the isotopologues show absorption lines of similar intensities at an isotopic abundance close to the natural one. This is done to avoid lines that saturate, which could deteriorate the data analysis, but also because most of IRMs find values that are close to the natural isotopic abundance. Note that other kind of restrictions could be considered~\cite{robinson-oe-2019}.
First, let us consider carbon dioxide and its $^{13}$C/$^{12}$C ratio.

\subsubsection{Carbon dioxide and $ \delta {}^{13}$C}
The ratio ${}^{13}r={}^{13}$C/$^{12}$C can be studied for carbon dioxide by analyzing the spectral region around 4.36µm, which mainly shows absorption lines coming from the $ \nu_3 $ band of $^{12}$CO$_2 $ and $^{13}$CO$_2 $ at similar intensities when these isotopologues are close to the natural isotopic abundance. In our case, we target the spectral region roughly between 4355nm and 4373nm.
Then, we fill a 10cm long cell with a gas mixture of carbon dioxide and synthetic air (the partial pressures will be detailed later) and we record the transmission spectrum of the sample. The experimental spectrum is then compared to a fitted spectrum in the same way as explained before, and with free parameters that are the ratios ${}^{13}r$, ${}^{18}r$, the partial pressure of carbon dioxide $p_{\text{CO}_2}$, a global frequency shift of the spectrum, and a function to correct the non-perfect baseline.

From the comparison between the experimental data and the fitted spectrum, one can extract the isotopic ratio $ {}^{13}r $ given by the least-squares regression. However, this way of doing will give us an absolute measurement, which is generally very complex to perform accurately since a lot of phenomena can induce systematic errors~\cite{sessions-jss-2006,yang-massspecrev-2009}.
Nevertheless, it is generally possible to bypass significantly this issue by measuring a relative isotopic ratio, rather than an absolute one. By using a reference sample that is pre-characterized in isotopic ratio $ {}^{13}r_\text{ref} $, one can measure the relative value between an unknown sample of isotopic ratio $ {}^{13}r_\text{exp} $ and the one that is pre-characterized. The relative value is usually expressed using the $ \delta $ notation in $ \permil $ which is, in the case of $^{13}$C and $^{12}$C, defined as:
\begin{equation}
\delta{}^{13}\text{C} = 1000 \left ( \frac{{}^{13}r_\text{exp}}{{}^{13}r_\text{ref}} -1 \right ) \permil \quad .
\end{equation}
In the case of the $ {}^{13}r $ ratio measurement, the reference sample is usually related to the Vienna Pee Dee Belemnite (VPDB)~\cite{iaea-book-1995}. Note that absolute measurements are possible, but they require much more complex setups, and are usually used for highly precise measurements~\cite{fleisher-natphys-2021}. To summarize, an isotope ratio spectrometer needs to be calibrated.

Several calibration procedures are reported in the literature and a practical guide can be found in Ref.~\cite{griffith-amt-2018}. For an ideal calibration of our setup, the procedure proposed in Ref.~\cite{griffith-amt-2018} should be followed, but in our case, we will choose a more simple approach. The reason for this is that for an ideal calibration procedure, the setup being calibrated should be sufficiently sensitive to the isotopic ratio ${}^{18} r$, which is not our case. Moreover, such kind of calibration procedures are more suitable when targeting high accuracy and low uncertainty measurements, which is not the purpose of this study. Hence we chose an approach inspired by Ref.~\cite{griffith-amt-2018} but more suitable for our setup.

As with other calibration procedures, ours consists of studying the response of the spectrometer with pre-chracterized isotopic ratio gas samples of CO$_2$. Here, we used samples originating from two commercially available gas bottles with a known but different $ \delta{}^{13}$C value to improve the consistency of our measurements~\cite{coplen-analchem-2006}. To confirm and refine the values given by the manufacturer, several samples from these two bottles were also characterized using an isotope ratio mass spectrometer (Delta V ThermoScientific) giving us the following values $ \delta{}^{13}$C~$=(-2.6 \pm 0.1 ) \permil$ and $ \delta{}^{13}$C~$=(-24.6 \pm 0.1 ) \permil$ respectively, relatively to the VPDB.

\begin{figure}[htb]
\centering
\includegraphics[]{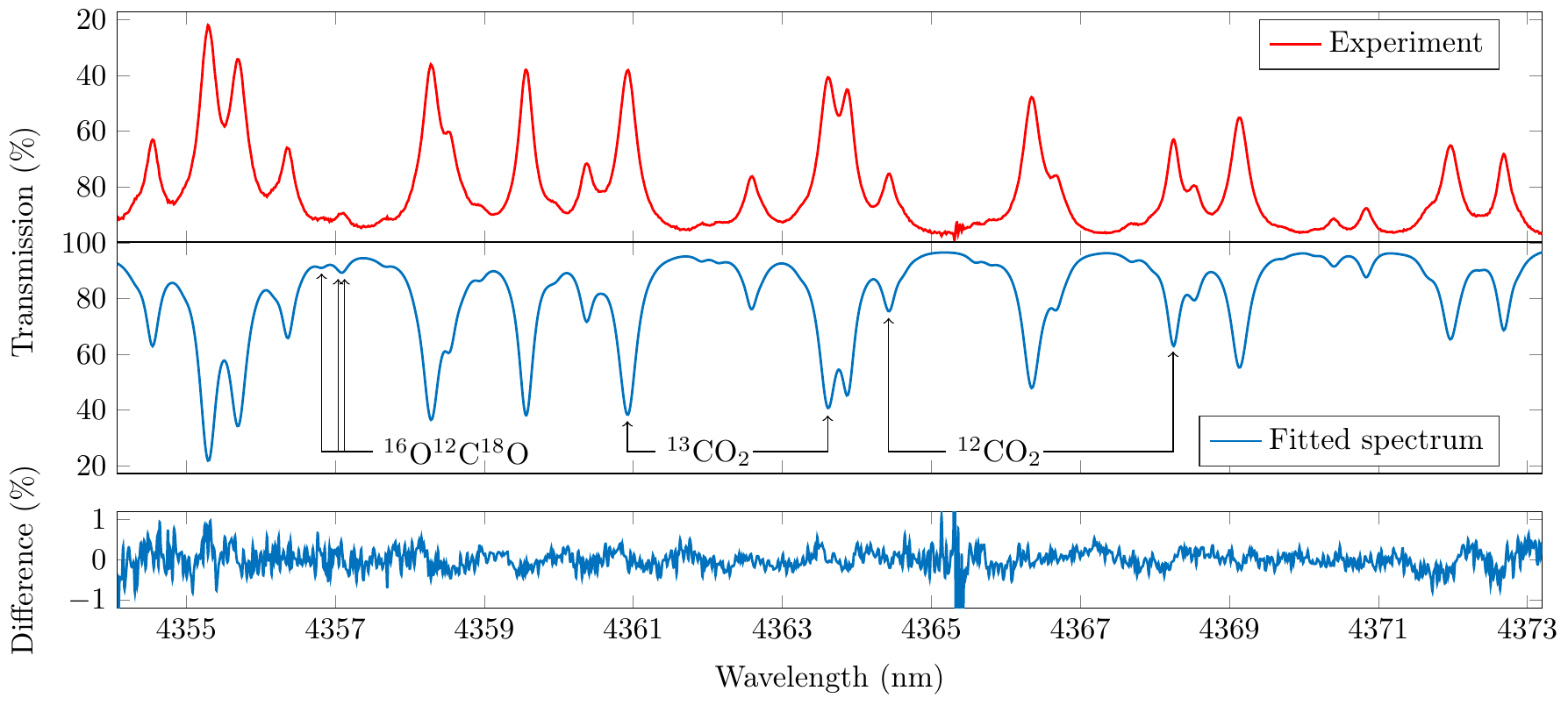}
\caption{(Top) Absorption spectrum of a 2.5\% CO$_2$ and 97.5\% air gas mixture recorded around 4364nm at a pressure of 800mbar using a 20s long interferogram. (Middle) Fitted absorption spectrum using line parameters coming from the HITRAN database. (Bottom) Difference between the experimental and fitted spectra.}
\label{fig:absco2unk}
\end{figure}

To perform the calibration with our setup, we first fill a 10cm long cell with a mixture of 2.5\% of CO$_2$ coming from one pre-characterized gas bottle and the rest of synthetic air, at a total pressure of 800mbar. Note that for the calibration procedure and the further measurements, we will work at the same partial pressure of CO$_2$ and at the same total pressure. This is done to avoid any potential pressure dependence of our measurements. One could also calibrate the setup in partial pressure, but this increases the complexity and induces a new potential source of errors.
Then, we record the absorption spectrum of the sample in the cell in the same way as detailed before using 20s long interferograms. The spectrum is then compared to a fitted spectrum and the $ {}^{13}r_\text{exp} $ ratio is extracted from the least-squares regression. A typical example of recorded spectrum is shown in \autoref{fig:absco2unk} along with the fitted spectrum used for comparison and the difference between the two sets of data. The same measurement under the same condition is made with the second pre-characterized gas bottle. As explained before, the obtained values can be very different from the expected ones, which is due to the non-perfect response of our setup. However, from the measurements with the two pre-characterized gas bottles, we now posses a way of linking the measured $ {}^{13}r_\text{exp} $ values with the expected ones using a calibration equation that we choose as linear.

\begin{figure}[htb]
\centering
\includegraphics[]{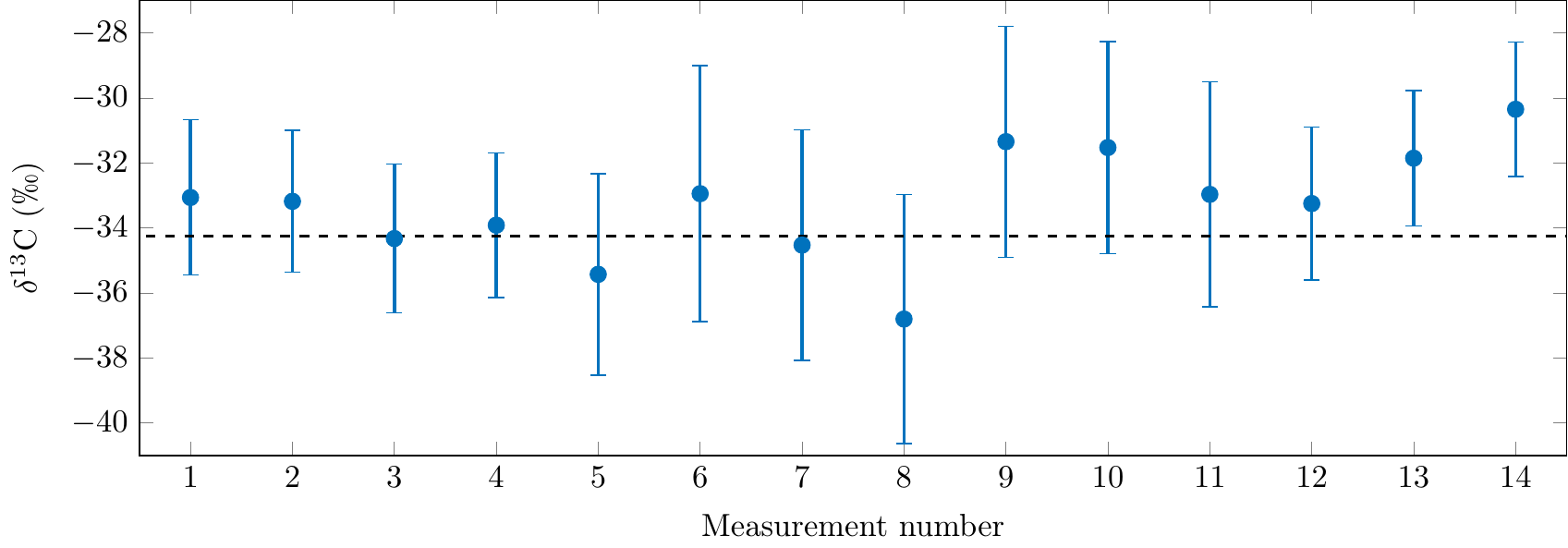}
\caption{Graph showing $ \delta{}^{13}$C measurements and their error bars at a confidence level of 95\% (2 standard deviations) of gas samples coming a gas bottle under characterization. The dashed line represents a $ \delta{}^{13}$C measurement of the same bottle obtained using an isotope ratio mass spectrometer.}
\label{fig:mesdelta}
\end{figure}

Once the calibration is done, we can investigate an unknown sample of CO$_2$ which will be here a gas bottle of unknown $ \delta{}^{13}$C value. In the same condition as in the calibration procedure, we record a transmission spectrum that is compared with a fitted spectrum.
From the regression, one can extract the isotopic ratio $ {}^{13}r_\text{exp} $ for this gas sample and by applying the inverted calibration equation, we can obtain a $ \delta{}^{13}$C measurement for this sample. The calibration procedure and the measurement using gas samples from the bottle under characterization is repeated several times, over several days. The obtained results and their associated error bars, which are given by two times the calibrated standard deviation resulting from the least-squares regression, are presented in \autoref{fig:mesdelta}.

A simple statistical analysis of the results obtained can give us the mean value and their dispersion which is given by the standard deviation of this whole set of measurements. We found a $ \delta{}^{13}$C value that is:
 \[ \delta{}^{13}\text{C} = (-33.2 \pm 1.7 ) \permil  \quad .\]
Note that for this measurement set, individual measurements show standard deviation close to 1.1\permil{} for half of them, and close to 1.8\permil{} for the other half. The reason for such differences might be due to the setup stability that can slightly change over the days. The value obtained from the statistical analysis can be compared with a set of measurements made by the isotope ratio mass spectrometer previously used. The result obtained with this instrument is $ \delta{}^{13}$C~$= (-34.3 \pm 0.1 ) \permil $, which shows a good agreement within errors with the value given by our isotope ratio dual-comb spectrometer.

\subsubsection{Nitrous oxide and $ \delta {}^{15}$N$^{\alpha}$}
Still using the frequency agility of the setup, it is possible to investigate nitrous oxide since absorption bands of this molecule and its isotopologues are present below 5µm~\cite{bailey-acs-2020}. In our case we study the spectral region around 4.63µm where the nitrous oxide isotopologues $^{14}$N$_2$O and $^{14}$N$^{15}$NO show absorption lines with similar intensities when close to the natural isotopic abundance.
Hence, in the same way as we did for carbon dioxide and the $ \delta {}^{13}$C measurement, it is possible to perform $ \delta {}^{15}$N$^\alpha$ measurements, the upperscript $ \alpha $ being used to differentiate $^{14}$N$^{15}$NO and $^{15}$N$^{14}$NO due to the fact that they do not show the same absorption features.
As shown with the measurement of $ \delta {}^{13}$C, a complete measurement of the $ \delta {}^{15}$N$^\alpha$ ratio would require a calibration of our spectrometer with pre-characterized samples of nitrous oxide in isotopic ratio. Considering we already presented the same study for the measurement of $ \delta {}^{13}$C, here we will only present raw uncalibrated measurements to illustrate the features of our setup regarding nitrous oxide studies.

\begin{figure}[htb]
\centering
\includegraphics[]{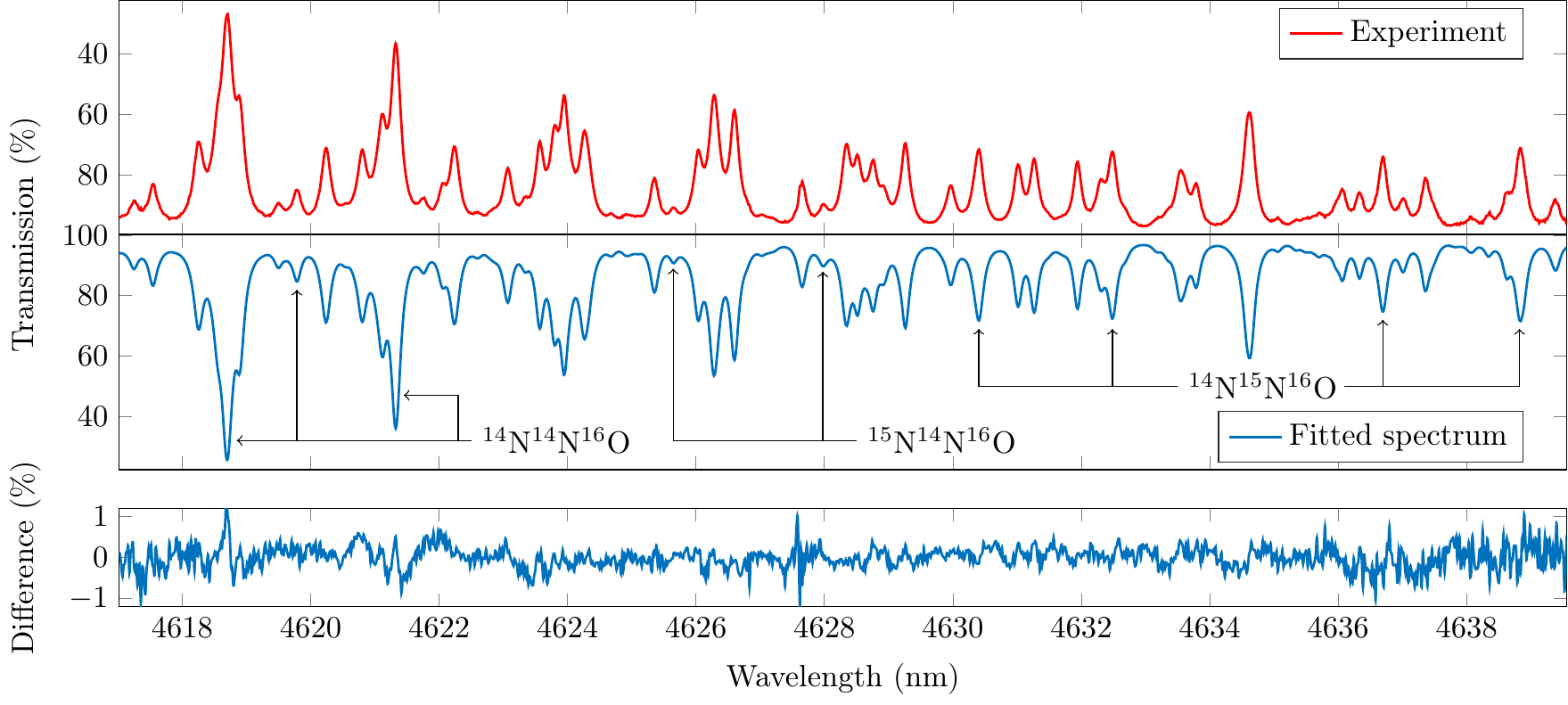}
\caption{(Top) Absorption spectrum of a 10\% N$_2$O and 90\% air gas mixture recorded around 4628nm at a total pressure of 455mbar using a 60s long interferogram. (Middle) Fitted absorption spectrum using line parameters coming from the HITRAN database. (Bottom) Difference between the experimental and fitted spectra.}
\label{fig:n2oabsspec}
\end{figure}

The setup is set with an idler CW laser at 2342nm, which gives a MIR signal that is centered at 4627nm. The 10cm cell is filled with a gas mixture of 10\% nitrous oxide and 90\% synthetic air at a total pressure of 455mbar. A 60s long interferogram is then recorded and treated as explained before to recover the absorption spectrum of the gas injected in the cell, and a typical example of obtained spectrum is shown is \autoref{fig:n2oabsspec}~(top). As with carbon dioxide, the spectrum is compared with a fitted spectrum obtained by least-squares regression which is shown in \autoref{fig:n2oabsspec}~(middle), and the difference between the two sets of data is presented in \autoref{fig:n2oabsspec}~(bottom). For the least-squares regression, the model used here is still a set a Voigt profile with N$_2$O line parameters given by the HTIRAN database, and only the free parameters have been changed to accommodate for the difference of study in isotopic ratios.
Several recordings showed a raw standard deviation resulting from the regression around 2\permil, which is slightly above the raw standard deviations obtained in the $ {}^{13}r_\text{exp} $ measurements and at the price of a longer interferogram recording. However, it is not possible to give any concrete signification to this value, especially on the measurement uncertainty for the $ \delta {}^{15}$N$^{\alpha}$ ratio as long as the setup is not calibrated yet.

\section{Discussions and conclusion}
In this article, we presented a free running dual-comb setup operating in the mid-infrared for isotope ratio measurements. We showed that by an appropriate choice of the analysis spectral range and by calibrating our setup against pre-characterized gas samples, it was possible to accurately measure the relative isotopic ratio $ \delta {}^{13}$C with a standard deviation below 2\permil. The measurements performed present a very good repeatability, and a comparison with data obtained using a mass spectrometer showed a very good agreement, thus validating the use of dual-comb spectroscopy for isotope ratio measurements.
The work presented here, and especially the measurements uncertainties, could be improved by stabilizing against frequency standards the CW lasers used, by decreasing the repetition rates of the combs for a higher spectral resolution, by increasing the spectral windows investigated with a shorter PPLN crystal, or even by using more advanced procedures for data analyses~\cite{castrillo-pra-2012}.
We believe these potential enhancements could improve the measurements delivered by our setup and thus compete with several isotope ratio spectrometers currently available on the market. However, let us note that the typical uncertainties obtained here can already be enough for several applications, such as for the detection of Helicobacter Pylori infection by breath test which requires a threshold around 3\permil{}~\cite{perets-2019}, and even comparable with the precision of embedded spectrometers on martian rovers like Curiosity which is around 2\permil{}~\cite{webster-pss-2011,mahaffy-ssr-2012}. 

Although any dual-comb spectrometer could be used for such isotope ratio measurements, electro-optic frequency combs seem to be very good candidates for miniaturizing such kind of setups since they can be generated on-chip~\cite{rueda-nature-2019,zhang-nature-2019}, which could enable easier on field isotope ratio measurements with embedded systems.
Regarding the applications, in the case of carbon isotope ratio measurements, changes in the absorption length could result in the study of atmospheric conditions, or using pure carbon dioxide samples. The first possibility has shown particular interests~\cite{nelson-apb-2008}, whereas the second one could avoid the use of diluting the samples in air and be particularly useful for applications where relatively high amount of CO$_2$ can be used.
The technique presented here could in principle also be used to target other molecular species, such as sulfur isotopes via the study of sulfur dioxide~\cite{christensen-acs-2007}, but this would require to go further in the MIR, above 5µm where PPLN crystals are no more transparent. However, other materials and techniques were reported to be able to study such spectral regions~\cite{luo-ol-2020,kowligy-ol-2020}. To conclude, the results presented here show that dual-comb spectroscopy has a high potential in the measurements of isotope ratios, and we believe that these results could paved the way to new kinds of isotope ratio spectrometers.

\section*{Acknowledgments}
The authors would like to thank Vincent Boudon for his availability to answer many questions related to the HITRAN database and spectroscopy, and Ivan Jovovic for his help in running the isotope ratio mass spectrometer measurements.

\section*{Funding}
Centre national de la recherche scientifique (CNRS), Conseil régional de Bourgogne Franche-Comté, iXCore Research Fondation, Agence Nationale de la Recherche (ANR-19-CE47-0008, ANR-15-IDEX-0003, ANR-21-CE42-0026-01, ANR-21-ESRE-0040).

\section*{Conflict of interest} The authors declare no conflict of interest.

\printbibliography

@Article{jaubert-nature-2016,
  author    = {Jacques Jaubert and Sophie Verheyden and Dominique Genty and Michel Soulier and Hai Cheng and Dominique Blamart and Christian Burlet and Hubert Camus and Serge Delaby and Damien Deldicque and R. Lawrence Edwards and Catherine Ferrier and Fran{\c{c}}ois Lacrampe-Cuyaub{\`{e}}re and Fran{\c{c}}ois L{\'{e}}v{\^{e}}que and Fr{\'{e}}d{\'{e}}ric Maksud and Pascal Mora and Xavier Muth and {\'{E}}douard R{\'{e}}gnier and Jean-Noël Rouzaud and Fr{\'{e}}d{\'{e}}ric Santos},
  title     = {Early Neanderthal constructions deep in Bruniquel Cave in southwestern France},
  journal   = {Nature},
  year      = {2016},
  volume    = {534},
  number    = {7605},
  pages     = {111--114},
  month     = may,
  doi       = {10.1038/nature18291},
  publisher = {Springer Science and Business Media {LLC}},
  url       = {https://doi.org/10.1038/nature18291},
}

@Article{webster-science-2013,
  author    = {Webster, Chris R. and Mahaffy, Paul R. and Flesch, Gregory J. and Niles, Paul B. and Jones, John H. and Leshin, Laurie A. and Atreya, Sushil K. and Stern, Jennifer C. and Christensen, Lance E. and Owen, Tobias and Franz, Heather and Pepin, Robert O. and Steele, Andrew and the MSL Science Team},
  title     = {Isotope Ratios of H, C, and O in CO$_2$ and H$_2$O of the Martian Atmosphere},
  journal   = {Science},
  year      = {2013},
  volume    = {341},
  number    = {6143},
  pages     = {260--263},
  issn      = {0036-8075},
  doi       = {10.1126/science.1237961},
  eprint    = {https://science.sciencemag.org/content/341/6143/260.full.pdf},
  publisher = {American Association for the Advancement of Science},
  url       = {https://science.sciencemag.org/content/341/6143/260},
}

@Article{tartese-pnas-2018,
  author    = {Tart{\`e}se, Romain and Chaussidon, Marc and Gurenko, Andrey and Delarue, Fr{\'e}d{\'e}ric and Robert, Fran{\c c}ois},
  title     = {Insights into the origin of carbonaceous chondrite organics from their triple oxygen isotope composition},
  journal   = {Proceedings of the National Academy of Sciences},
  year      = {2018},
  volume    = {115},
  number    = {34},
  pages     = {8535--8540},
  issn      = {0027-8424},
  doi       = {10.1073/pnas.1808101115},
  eprint    = {https://www.pnas.org/content/115/34/8535.full.pdf},
  publisher = {National Academy of Sciences},
  url       = {https://www.pnas.org/content/115/34/8535},
}

@Article{ruf-pnas-2017,
  author    = {Ruf, Alexander and Kanawati, Basem and Hertkorn, Norbert and Yin, Qing-Zhu and Moritz, Franco and Harir, Mourad and Lucio, Marianna and Michalke, Bernhard and Wimpenny, Joshua and Shilobreeva, Svetlana and Bronsky, Basil and Saraykin, Vladimir and Gabelica, Zelimir and Gougeon, R{\'e}gis D. and Quirico, Eric and Ralew, Stefan and Jakubowski, Tomasz and Haack, Henning and Gonsior, Michael and Jenniskens, Peter and Hinman, Nancy W. and Schmitt-Kopplin, Philippe},
  title     = {Previously unknown class of metalorganic compounds revealed in meteorites},
  journal   = {Proceedings of the National Academy of Sciences},
  year      = {2017},
  volume    = {114},
  number    = {11},
  pages     = {2819--2824},
  issn      = {0027-8424},
  doi       = {10.1073/pnas.1616019114},
  eprint    = {https://www.pnas.org/content/114/11/2819.full.pdf},
  publisher = {National Academy of Sciences},
  url       = {https://www.pnas.org/content/114/11/2819},
}

@Article{brooks-jafc-2002,
  author    = {J. Ren{\'{e}}e Brooks and Nina Buchmann and Sue Phillips and Bruce Ehleringer and R. David Evans and Mike Lott and Luiz A. Martinelli and William T. Pockman and Darren Sandquist and Jed P. Sparks and Lynda Sperry and Dave Williams and James R. Ehleringer},
  title     = {Heavy and Light Beer:~ A Carbon Isotope Approach To Detect C4Carbon in Beers of Different Origins, Styles, and Prices},
  journal   = {Journal of Agricultural and Food Chemistry},
  year      = {2002},
  volume    = {50},
  number    = {22},
  pages     = {6413--6418},
  month     = oct,
  doi       = {10.1021/jf020594k},
  publisher = {American Chemical Society ({ACS})},
  url       = {https://doi.org/10.1021/jf020594k},
}

@Article{frei-scirep-2015,
  author    = {Karin Margarita Frei and Ulla Mannering and Kristian Kristiansen and Morten E. Allentoft and Andrew S. Wilson and Irene Skals and Silvana Tridico and Marie Louise Nosch and Eske Willerslev and Leon Clarke and Robert Frei},
  title     = {Tracing the dynamic life story of a Bronze Age Female},
  journal   = {Scientific Reports},
  year      = {2015},
  volume    = {5},
  number    = {1},
  month     = may,
  doi       = {10.1038/srep10431},
  publisher = {Springer Science and Business Media {LLC}},
  url       = {https://doi.org/10.1038/srep10431},
}

@Article{magdas-iehs-2012,
  author    = {Dana Alina Magdas and Stela Cuna and Gabriela Cristea and Roxana Elena Ionete and Diana Costinel},
  title     = {Stable isotopes determination in some Romanian wines},
  journal   = {Isotopes in Environmental and Health Studies},
  year      = {2012},
  volume    = {48},
  number    = {2},
  pages     = {345-353},
  note      = {PMID: 22397311},
  doi       = {10.1080/10256016.2012.661731},
  eprint    = {https://doi.org/10.1080/10256016.2012.661731},
  publisher = {Taylor \& Francis},
  url       = {https://doi.org/10.1080/10256016.2012.661731},
}

@Article{day-jsfa-1995,
  author   = {Day, Martin P and Zhang, Benli and Martin, Gerard J},
  title    = {Determination of the geographical origin of wine using joint analysis of elemental and isotopic composition. II—Differentiation of the principal production zones in france for the 1990 vintage},
  journal  = {Journal of the Science of Food and Agriculture},
  year     = {1995},
  volume   = {67},
  number   = {1},
  pages    = {113-123},
  doi      = {10.1002/jsfa.2740670118},
  eprint   = {https://onlinelibrary.wiley.com/doi/pdf/10.1002/jsfa.2740670118},
  url      = {https://onlinelibrary.wiley.com/doi/abs/10.1002/jsfa.2740670118},
}

@Article{parriaux-aop-2020,
  author    = {Alexandre Parriaux and Kamal Hammani and Guy Millot},
  title     = {Electro-optic frequency combs},
  journal   = {Adv. Opt. Photon.},
  year      = {2020},
  volume    = {12},
  number    = {1},
  pages     = {223--287},
  month     = {Mar},
  doi       = {10.1364/AOP.382052},
  publisher = {OSA},
  url       = {http://aop.osa.org/abstract.cfm?URI=aop-12-1-223},
}

@Article{yan-lsa-2017,
  author    = {Yan, Ming and Luo, Pei-Ling and Iwakuni, Kana and Millot, Guy and H{\"a}nsch, Theodor W. and Picqu{\'e}, Nathalie},
  title     = {Mid-infrared dual-comb spectroscopy with electro-optic modulators},
  journal   = {Light: Science \& Applications},
  year      = {2017},
  volume    = {6},
  pages     = {e17076},
  month     = {Oct},
  day       = {20},
  publisher = {The Author(s)},
  url       = {https://doi.org/10.1038/lsa.2017.76},
}

@Article{luo-pccp-2019,
  author    = {Pei-Ling Luo and Er-Chien Horng and Yu-Chan Guan},
  title     = {Fast molecular fingerprinting with a coherent, rapidly tunable dual-comb spectrometer near 3 $\mu$m},
  journal   = {Physical Chemistry Chemical Physics},
  year      = {2019},
  volume    = {21},
  number    = {33},
  pages     = {18400--18405},
  doi       = {10.1039/c9cp03090e},
  publisher = {Royal Society of Chemistry ({RSC})},
  url       = {https://doi.org/10.1039/c9cp03090e},
}

@Article{jerez-acsphot-2018,
  author  = {Jerez, Borja and Martín-Mateos, Pedro and Walla, Frederik and de Dios, Cristina and Acedo, Pablo},
  title   = {Flexible Electro-Optic, Single-Crystal Difference Frequency Generation Architecture for Ultrafast Mid-Infrared Dual-Comb Spectroscopy},
  journal = {ACS Photonics},
  year    = {2018},
  volume  = {5},
  number  = {6},
  pages   = {2348-2353},
  doi     = {10.1021/acsphotonics.8b00143},
  eprint  = {https://doi.org/10.1021/acsphotonics.8b00143},
  url     = {https://doi.org/10.1021/acsphotonics.8b00143},
}

@Article{millot-natphot-2016,
  author    = {Millot, Guy and Pitois, Stéphane and Yan, Ming and Hovhannisyan, Tatevik and Bendahmane, Abdelkrim and Hänsch, Theodor W. and Picqué, Nathalie},
  title     = {Frequency-agile dual-comb spectroscopy},
  journal   = {Nature Photonics},
  year      = {2016},
  volume    = {10},
  pages     = {27},
  publisher = {Nature Publishing Group},
  url       = {http://dx.doi.org/10.1038/nphoton.2015.250},
}

@Article{coddington-optica-2016,
  author    = {Ian Coddington and Nathan Newbury and William Swann},
  title     = {Dual-comb spectroscopy},
  journal   = {Optica},
  year      = {2016},
  volume    = {3},
  number    = {4},
  pages     = {414--426},
  month     = {Apr},
  doi       = {10.1364/OPTICA.3.000414},
  publisher = {OSA},
  url       = {http://www.osapublishing.org/optica/abstract.cfm?URI=optica-3-4-414},
}

@Article{parriaux-ol-2017,
  author    = {A. Parriaux and M. Conforti and A. Bendahmane and J. Fatome and C. Finot and S. Trillo and N. Picqu\'{e} and G. Millot},
  title     = {Spectral broadening of picosecond pulses forming dispersive shock waves in optical fibers},
  journal   = {Opt. Lett.},
  year      = {2017},
  volume    = {42},
  number    = {15},
  pages     = {3044--3047},
  month     = {Aug},
  doi       = {10.1364/OL.42.003044},
  publisher = {OSA},
  url       = {http://ol.osa.org/abstract.cfm?URI=ol-42-15-3044},
}

@Article{robinson-oe-2019,
  author    = {Iain Robinson and Helen L. Butcher and Neil A. Macleod and Damien Weidmann},
  title     = {Hollow waveguide integrated laser spectrometer for $^{12}$CO$_2$/$^{12}$CO$_2$ analysis},
  journal   = {Opt. Express},
  year      = {2019},
  volume    = {27},
  number    = {24},
  pages     = {35670--35688},
  month     = {Nov},
  doi       = {10.1364/OE.27.035670},
  publisher = {OSA},
  url       = {http://www.opticsexpress.org/abstract.cfm?URI=oe-27-24-35670},
}

@Article{camin-revfood-2016,
  author   = {Camin, Federica and Bontempo, Luana and Perini, Matteo and Piasentier, Edi},
  title    = {Stable Isotope Ratio Analysis for Assessing the Authenticity of Food of Animal Origin},
  journal  = {Comprehensive Reviews in Food Science and Food Safety},
  year     = {2016},
  volume   = {15},
  number   = {5},
  pages    = {868-877},
  doi      = {10.1111/1541-4337.12219},
  eprint   = {https://onlinelibrary.wiley.com/doi/pdf/10.1111/1541-4337.12219},
  url      = {https://onlinelibrary.wiley.com/doi/abs/10.1111/1541-4337.12219},
}

@Article{chung-foodchem-2018,
  author   = {Ill-Min Chung and Jae-Gu Han and Won-Sik Kong and Jae-Kwang Kim and Min-Jeong An and Ji-Hee Lee and Yeon-Ju An and Mun Yhung Jung and Seung-Hyun Kim},
  title    = {Regional discrimination of Agaricus bisporus mushroom using the natural stable isotope ratios},
  journal  = {Food Chemistry},
  year     = {2018},
  volume   = {264},
  pages    = {92 - 100},
  issn     = {0308-8146},
  doi      = {https://doi.org/10.1016/j.foodchem.2018.04.138},
  url      = {http://www.sciencedirect.com/science/article/pii/S0308814618307878},
}

@Article{camin-foodchem-2004,
  author    = {Federica Camin and Karine Wietzerbin and Anaisabel Blanch Cortes and Georg Haberhauer and Mich{\'{e}}le Lees and Giuseppe Versini},
  title     = {Application of Multielement Stable Isotope Ratio Analysis to the Characterization of French, Italian, and Spanish Cheeses},
  journal   = {Journal of Agricultural and Food Chemistry},
  year      = {2004},
  volume    = {52},
  number    = {21},
  pages     = {6592--6601},
  month     = oct,
  doi       = {10.1021/jf040062z},
  publisher = {American Chemical Society ({ACS})},
  url       = {https://doi.org/10.1021/jf040062z},
}

@Book{iaea-book-1995,
  title     = {Reference and Intercomparison Materials for Stable Isotopes of Light Elements},
  publisher = {INTERNATIONAL ATOMIC ENERGY AGENCY},
  year      = {1995},
  number    = {825},
  series    = {TECDOC Series},
  address   = {Vienna},
  url       = {https://www.iaea.org/publications/5471/reference-and-intercomparison-materials-for-stable-isotopes-of-light-elements},
}

@Article{graham-1987,
  author    = {David Y. Graham and Doyle J. Evans and Lesley C. Alpert and Peter D. Klein and Dolores G. Evans and Antone R. Opekun and Thomas W. Boutton},
  title     = {Campylobacter pylori detected noninvasively by the $^{13}$C-urea breath test},
  journal   = {The Lancet},
  year      = {1987},
  volume    = {329},
  number    = {8543},
  pages     = {1174--1177},
  month     = may,
  doi       = {10.1016/s0140-6736(87)92145-3},
  publisher = {Elsevier {BV}},
  url       = {https://doi.org/10.1016/s0140-6736(87)92145-3},
}

@Article{perets-2019,
  author   = {Perets, Tsachi Tsadok and Gingold-Belfer, Rachel and Leibovitzh, Haim and Itskoviz, David and Schmilovitz-Weiss, Hemda and Snir, Yifat and Dickman, Ram and Dotan, Iris and Levi, Zohar and Boltin, Doron},
  title    = {Optimization of 13C-urea breath test threshold levels for the detection of Helicobacter pylori infection in a national referral laboratory},
  journal  = {Journal of Clinical Laboratory Analysis},
  year     = {2019},
  volume   = {33},
  number   = {2},
  pages    = {e22674},
  doi      = {10.1002/jcla.22674},
  eprint   = {https://onlinelibrary.wiley.com/doi/pdf/10.1002/jcla.22674},
  url      = {https://onlinelibrary.wiley.com/doi/abs/10.1002/jcla.22674},
}

@Misc{hitran,
  author = {\url{http://www.cfa.harvard.edu/hitran/}},
}

@article{hitran2020,
title = {The HITRAN2020 molecular spectroscopic database},
journal = {Journal of Quantitative Spectroscopy and Radiative Transfer},
pages = {107949},
year = {2021},
issn = {0022-4073},
doi = {https://doi.org/10.1016/j.jqsrt.2021.107949},
url = {https://www.sciencedirect.com/science/article/pii/S0022407321004416},
author = {I.E. Gordon and L.S. Rothman and R.J. Hargreaves and R. Hashemi and E.V. Karlovets and F.M. Skinner and E.K. Conway and C. Hill and R.V. Kochanov and Y. Tan and P. Wcisło and A.A. Finenko and K. Nelson and P.F. Bernath and M. Birk and V. Boudon and A. Campargue and K.V. Chance and A. Coustenis and B.J. Drouin and J.–M. Flaud and R.R. Gamache and J.T. Hodges and D. Jacquemart and E.J. Mlawer and A.V. Nikitin and V.I. Perevalov and M. Rotger and J. Tennyson and G.C. Toon and H. Tran and V.G. Tyuterev and E.M. Adkins and A. Baker and A. Barbe and E. Canè and A.G. Császár and A. Dudaryonok and O. Egorov and A.J. Fleisher and H. Fleurbaey and A. Foltynowicz and T. Furtenbacher and J.J. Harrison and J.–M. Hartmann and V.–M. Horneman and X. Huang and T. Karman and J. Karns and S. Kassi and I. Kleiner and V. Kofman and F. Kwabia–Tchana and N.N. Lavrentieva and T.J. Lee and D.A. Long and A.A. Lukashevskaya and O.M. Lyulin and V.Yu. Makhnev and W. Matt and S.T. Massie and M. Melosso and S.N. Mikhailenko and D. Mondelain and H.S.P. Müller and O.V. Naumenko and A. Perrin and O.L. Polyansky and E. Raddaoui and P.L. Raston and Z.D. Reed and M. Rey and C. Richard and R. Tóbiás and I. Sadiek and D.W. Schwenke and E. Starikova and K. Sung and F. Tamassia and S.A. Tashkun and J. Vander Auwera and I.A. Vasilenko and A.A. Vigasin and G.L. Villanueva and B. Vispoel and G. Wagner and A. Yachmenev and S.N. Yurchenko}
}

@Article{griffith-amt-2018,
AUTHOR = {Griffith, D. W. T.},
TITLE = {Calibration of isotopologue-specific optical trace gas analysers: a practical guide},
JOURNAL = {Atmospheric Measurement Techniques},
VOLUME = {11},
YEAR = {2018},
NUMBER = {11},
PAGES = {6189--6201},
URL = {https://amt.copernicus.org/articles/11/6189/2018/},
DOI = {10.5194/amt-11-6189-2018}
}

@article{fleisher-natphys-2021,
  doi = {10.1038/s41567-021-01226-y},
  url = {https://doi.org/10.1038/s41567-021-01226-y},
  year = {2021},
  month = apr,
  publisher = {Springer Science and Business Media {LLC}},
  author = {Adam J. Fleisher and Hongming Yi and Abneesh Srivastava and Oleg L. Polyansky and Nikolai F. Zobov and Joseph T. Hodges},
  title = {Absolute 13C/12C isotope amount ratio for Vienna {PeeDee} Belemnite from infrared absorption spectroscopy},
  journal = {Nature Physics}
}

@article{bailey-acs-2020,
author = {Bailey, D. Michelle and Zhao, Gang and Fleisher, Adam J.},
title = {Precision Spectroscopy of Nitrous Oxide Isotopocules with a Cross-Dispersed Spectrometer and a Mid-Infrared Frequency Comb},
journal = {Analytical Chemistry},
volume = {92},
number = {20},
pages = {13759-13766},
year = {2020},
doi = {10.1021/acs.analchem.0c01868},
note ={PMID: 32942855},
URL = {https://doi.org/10.1021/acs.analchem.0c01868},
eprint = {https://doi.org/10.1021/acs.analchem.0c01868}
}

@InBook{vodopyanov-wiley-2020,
  pages     = {1-11},
  title     = {Isotopologues Detection and Quantitative Analysis by Mid-Infrared Dual-Comb Laser Spectroscopy},
  publisher = {American Cancer Society},
  year      = {2020},
  author    = {Vodopyanov, Konstantin L.},
  isbn      = {9780470027318},
  booktitle = {Encyclopedia of Analytical Chemistry},
  doi       = {https://doi.org/10.1002/9780470027318.a9321},
  eprint    = {https://onlinelibrary.wiley.com/doi/pdf/10.1002/9780470027318.a9321},
  url       = {https://onlinelibrary.wiley.com/doi/abs/10.1002/9780470027318.a9321},
}

@Article{muraviev-natphot-2018,
  author    = {A. V. Muraviev and V. O. Smolski and Z. E. Loparo and K. L. Vodopyanov},
  title     = {Massively parallel sensing of trace molecules and their isotopologues with broadband subharmonic mid-infrared frequency combs},
  journal   = {Nature Photonics},
  year      = {2018},
  volume    = {12},
  number    = {4},
  pages     = {209--214},
  month     = mar,
  doi       = {10.1038/s41566-018-0135-2},
  publisher = {Springer Science and Business Media {LLC}},
  url       = {https://doi.org/10.1038/s41566-018-0135-2},
}

@Article{tian-pnas-2021,
  author       = {Tian, Zhen and Magna, Tom{\'a}{\v s} and Day, James M. D. and Mezger, Klaus and Scherer, Erik E. and Lodders, Katharina and Hin, Remco C. and Koefoed, Piers and Bloom, Hannah and Wang, Kun},
  title        = {Potassium isotope composition of Mars reveals a mechanism of planetary volatile retention},
  journal      = {Proceedings of the National Academy of Sciences},
  year         = {2021},
  volume       = {118},
  number       = {39},
  issn         = {0027-8424},
  doi          = {10.1073/pnas.2101155118},
  elocation-id = {e2101155118},
  eprint       = {https://www.pnas.org/content/118/39/e2101155118.full.pdf},
  publisher    = {National Academy of Sciences},
  url          = {https://www.pnas.org/content/118/39/e2101155118},
}

@InCollection{kerstel-book-2004,
  author    = {Erik Kerstel},
  title     = {Chapter 34 - Isotope Ratio Infrared Spectrometry},
  booktitle = {Handbook of Stable Isotope Analytical Techniques},
  publisher = {Elsevier},
  year      = {2004},
  editor    = {Pier A. {de Groot}},
  pages     = {759-787},
  address   = {Amsterdam},
  isbn      = {978-0-444-51114-0},
  doi       = {https://doi.org/10.1016/B978-044451114-0/50036-3},
  url       = {https://www.sciencedirect.com/science/article/pii/B9780444511140500363},
}

@Article{castrillo-pra-2012,
  author    = {Castrillo, Antonio and Dinesan, Hemanth and Casa, Giovanni and Galzerano, Gianluca and Laporta, Paolo and Gianfrani, Livio},
  title     = {Amount-ratio determinations of water isotopologues by dual-laser absorption spectrometry},
  journal   = {Phys. Rev. A},
  year      = {2012},
  volume    = {86},
  pages     = {052515},
  month     = {Nov},
  doi       = {10.1103/PhysRevA.86.052515},
  issue     = {5},
  numpages  = {7},
  publisher = {American Physical Society},
  url       = {https://link.aps.org/doi/10.1103/PhysRevA.86.052515},
}

@Article{luo-ol-2020,
  author    = {Pei-Ling Luo},
  title     = {Long-wave mid-infrared time-resolved dual-comb spectroscopy of short-lived intermediates},
  journal   = {Opt. Lett.},
  year      = {2020},
  volume    = {45},
  number    = {24},
  pages     = {6791--6794},
  month     = {Dec},
  doi       = {10.1364/OL.413754},
  publisher = {OSA},
  url       = {http://www.osapublishing.org/ol/abstract.cfm?URI=ol-45-24-6791},
}

@Article{kowligy-ol-2020,
  author    = {Abijith S. Kowligy and David R. Carlson and Daniel D. Hickstein and Henry Timmers and Alexander J. Lind and Peter G. Schunemann and Scott B. Papp and Scott A. Diddams},
  title     = {Mid-infrared frequency combs at 10 GHz},
  journal   = {Opt. Lett.},
  year      = {2020},
  volume    = {45},
  number    = {13},
  pages     = {3677--3680},
  month     = {Jul},
  doi       = {10.1364/OL.391651},
  publisher = {OSA},
  url       = {http://www.osapublishing.org/ol/abstract.cfm?URI=ol-45-13-3677},
}

@Article{christensen-acs-2007,
  author  = {Christensen, Lance E. and Brunner, Benjamin and Truong, Kasey N. and Mielke, Randall E. and Webster, Christopher R. and Coleman, Max},
  title   = {Measurement of Sulfur Isotope Compositions by Tunable Laser Spectroscopy of SO2},
  journal = {Analytical Chemistry},
  year    = {2007},
  volume  = {79},
  number  = {24},
  pages   = {9261-9268},
  note    = {PMID: 18020312},
  doi     = {10.1021/ac071040p},
  eprint  = {https://doi.org/10.1021/ac071040p},
  url     = {https://doi.org/10.1021/ac071040p},
}

@Article{nelson-apb-2008,
  author    = {D.D. Nelson and J.B. McManus and S.C. Herndon and M.S. Zahniser and B. Tuzson and L. Emmenegger},
  title     = {New method for isotopic ratio measurements of atmospheric carbon dioxide using a 4.3$\mu$m pulsed quantum cascade laser},
  journal   = {Applied Physics B},
  year      = {2008},
  volume    = {90},
  number    = {2},
  pages     = {301--309},
  month     = jan,
  doi       = {10.1007/s00340-007-2894-1},
  publisher = {Springer Science and Business Media {LLC}},
  url       = {https://doi.org/10.1007/s00340-007-2894-1},
}

@Article{coplen-analchem-2006,
  author  = {Coplen, Tyler B. and Brand, Willi A. and Gehre, Matthias and Gröning, Manfred and Meijer, Harro A. J. and Toman, Blaza and Verkouteren, R. Michael},
  title   = {New Guidelines for $\delta$13C Measurements},
  journal = {Analytical Chemistry},
  year    = {2006},
  volume  = {78},
  number  = {7},
  pages   = {2439-2441},
  note    = {PMID: 16579631},
  doi     = {10.1021/ac052027c},
  eprint  = {https://doi.org/10.1021/ac052027c},
  url     = { 
        https://doi.org/10.1021/ac052027c
    
},
}

@Article{sessions-jss-2006,
  author   = {Sessions, Alex L.},
  title    = {Isotope-ratio detection for gas chromatography},
  journal  = {Journal of Separation Science},
  year     = {2006},
  volume   = {29},
  number   = {12},
  pages    = {1946-1961},
  doi      = {https://doi.org/10.1002/jssc.200600002},
  eprint   = {https://analyticalsciencejournals.onlinelibrary.wiley.com/doi/pdf/10.1002/jssc.200600002},
  url      = {https://analyticalsciencejournals.onlinelibrary.wiley.com/doi/abs/10.1002/jssc.200600002},
}

@Article{becker-appopt-1992,
  author    = {Joseph F. Becker and Todd B. Sauke and Max Loewenstein},
  title     = {Stable isotope analysis using tunable diode laser spectroscopy},
  journal   = {Appl. Opt.},
  year      = {1992},
  volume    = {31},
  number    = {12},
  pages     = {1921--1927},
  month     = {Apr},
  doi       = {10.1364/AO.31.001921},
  publisher = {OSA},
  url       = {http://www.osapublishing.org/ao/abstract.cfm?URI=ao-31-12-1921},
}

@Article{yang-massspecrev-2009,
  author   = {Yang, Lu},
  title    = {Accurate and precise determination of isotopic ratios by MC-ICP-MS: A review},
  journal  = {Mass Spectrometry Reviews},
  year     = {2009},
  volume   = {28},
  number   = {6},
  pages    = {990-1011},
  doi      = {https://doi.org/10.1002/mas.20251},
  eprint   = {https://analyticalsciencejournals.onlinelibrary.wiley.com/doi/pdf/10.1002/mas.20251},
  url      = {https://analyticalsciencejournals.onlinelibrary.wiley.com/doi/abs/10.1002/mas.20251},
}

@Article{wahl-isoenvhealth-2006,
  author    = {Ed H. Wahl and Bernard Fidric and Chris W. Rella and Sergei Koulikov and Boris Kharlamov and Sze Tan and Alexander A. Kachanov and Bruce A. Richman and Eric R. Crosson and Barbara A. Paldus and Shashi Kalaskar and David R. Bowling},
  title     = {Applications of cavity ring-down spectroscopy to high precision isotope ratio measurement of 13C/12C in carbon dioxide},
  journal   = {Isotopes in Environmental and Health Studies},
  year      = {2006},
  volume    = {42},
  number    = {1},
  pages     = {21-35},
  note      = {PMID: 16500752},
  doi       = {10.1080/10256010500502934},
  eprint    = {https://doi.org/10.1080/10256010500502934},
  publisher = {Taylor \& Francis},
  url       = { 
        https://doi.org/10.1080/10256010500502934
    
},
}

@Article{waechter-apb-2007,
  author    = {H. Waechter and M.W. Sigrist},
  title     = {Mid-infrared laser spectroscopic determination of isotope ratios of N2O at trace levels using wavelength modulation and balanced path length detection},
  journal   = {Applied Physics B},
  year      = {2007},
  volume    = {87},
  number    = {3},
  pages     = {539--546},
  month     = feb,
  doi       = {10.1007/s00340-007-2576-z},
  publisher = {Springer Science and Business Media {LLC}},
  url       = {https://doi.org/10.1007/s00340-007-2576-z},
}

@Article{rueda-nature-2019,
  author    = {Alfredo Rueda and Florian Sedlmeir and Madhuri Kumari and Gerd Leuchs and Harald G. L. Schwefel},
  title     = {Resonant electro-optic frequency comb},
  journal   = {Nature},
  year      = {2019},
  volume    = {568},
  number    = {7752},
  pages     = {378--381},
  month     = apr,
  doi       = {10.1038/s41586-019-1110-x},
  publisher = {Springer Nature},
  url       = {https://doi.org/10.1038/s41586-019-1110-x},
}

@Article{zhang-nature-2019,
  author   = {Zhang, Mian and Buscaino, Brandon and Wang, Cheng and Shams-Ansari, Amirhassan and Reimer, Christian and Zhu, Rongrong and Kahn, Joseph M. and Loncar, Marko},
  title    = {Broadband electro-optic frequency comb generation in a lithium niobate microring resonator},
  journal  = {Nature},
  year     = {2019},
  issn     = {1476-4687},
  doi      = {10.1038/s41586-019-1008-7},
  url      = {https://doi.org/10.1038/s41586-019-1008-7},
}

@Article{webster-pss-2011,
  author   = {Christopher R. Webster and Paul R. Mahaffy},
  title    = {Determining the local abundance of Martian methane and its’ 13C/12C and D/H isotopic ratios for comparison with related gas and soil analysis on the 2011 Mars Science Laboratory (MSL) mission},
  journal  = {Planetary and Space Science},
  year     = {2011},
  volume   = {59},
  number   = {2},
  pages    = {271-283},
  issn     = {0032-0633},
  note     = {Methane on Mars: Current Observations, Interpretation and Future Plans},
  doi      = {https://doi.org/10.1016/j.pss.2010.08.021},
  url      = {https://www.sciencedirect.com/science/article/pii/S003206331000259X},
}

@Article{mahaffy-ssr-2012,
  author    = {Paul R. Mahaffy and Christopher R. Webster and Michel Cabane and Pamela G. Conrad and Patrice Coll and Sushil K. Atreya and Robert Arvey and Michael Barciniak and Mehdi Benna and Lora Bleacher and William B. Brinckerhoff and Jennifer L. Eigenbrode and Daniel Carignan and Mark Cascia and Robert A. Chalmers and Jason P. Dworkin and Therese Errigo and Paula Everson and Heather Franz and Rodger Farley and Steven Feng and Gregory Frazier and Caroline Freissinet and Daniel P. Glavin and Daniel N. Harpold and Douglas Hawk and Vincent Holmes and Christopher S. Johnson and Andrea Jones and Patrick Jordan and James Kellogg and Jesse Lewis and Eric Lyness and Charles A. Malespin and David K. Martin and John Maurer and Amy C. McAdam and Douglas McLennan and Thomas J. Nolan and Marvin Noriega and Alexander A. Pavlov and Benito Prats and Eric Raaen and Oren Sheinman and David Sheppard and James Smith and Jennifer C. Stern and Florence Tan and Melissa Trainer and Douglas W. Ming and Richard V. Morris and John Jones and Cindy Gundersen and Andrew Steele and James Wray and Oliver Botta and Laurie A. Leshin and Tobias Owen and Steve Battel and Bruce M. Jakosky and Heidi Manning and Steven Squyres and Rafael Navarro-Gonz{\'{a}}lez and Christopher P. McKay and Francois Raulin and Robert Sternberg and Arnaud Buch and Paul Sorensen and Robert Kline-Schoder and David Coscia and Cyril Szopa and Samuel Teinturier and Curt Baffes and Jason Feldman and Greg Flesch and Siamak Forouhar and Ray Garcia and Didier Keymeulen and Steve Woodward and Bruce P. Block and Ken Arnett and Ryan Miller and Charles Edmonson and Stephen Gorevan and Erik Mumm},
  title     = {The Sample Analysis at Mars Investigation and Instrument Suite},
  journal   = {Space Science Reviews},
  year      = {2012},
  volume    = {170},
  number    = {1-4},
  pages     = {401--478},
  month     = apr,
  doi       = {10.1007/s11214-012-9879-z},
  publisher = {Springer Science and Business Media {LLC}},
  url       = {https://doi.org/10.1007/s11214-012-9879-z},
}

@Article{house-pnas-2022,
  author       = {House, Christopher H. and Wong, Gregory M. and Webster, Christopher R. and Flesch, Gregory J. and Franz, Heather B. and Stern, Jennifer C. and Pavlov, Alex and Atreya, Sushil K. and Eigenbrode, Jennifer L. and Gilbert, Alexis and Hofmann, Amy E. and Millan, Ma{\"e}va and Steele, Andrew and Glavin, Daniel P. and Malespin, Charles A. and Mahaffy, Paul R.},
  title        = {Depleted carbon isotope compositions observed at Gale crater, Mars},
  journal      = {Proceedings of the National Academy of Sciences},
  year         = {2022},
  volume       = {119},
  number       = {4},
  issn         = {0027-8424},
  doi          = {10.1073/pnas.2115651119},
  elocation-id = {e2115651119},
  eprint       = {https://www.pnas.org/content/119/4/e2115651119.full.pdf},
  publisher    = {National Academy of Sciences},
  url          = {https://www.pnas.org/content/119/4/e2115651119},
}

@Article{picque-natphot-2019,
  author    = {Nathalie Picqu{\'{e}} and Theodor W. H\"{a}nsch},
  title     = {Frequency comb spectroscopy},
  journal   = {Nature Photonics},
  year      = {2019},
  volume    = {13},
  number    = {3},
  pages     = {146--157},
  month     = feb,
  doi       = {10.1038/s41566-018-0347-5},
  publisher = {Springer Science and Business Media {LLC}},
  url       = {https://doi.org/10.1038/s41566-018-0347-5},
}

@Article{fortier-commphys-2019,
  author    = {Tara Fortier and Esther Baumann},
  title     = {20 years of developments in optical frequency comb technology and applications},
  journal   = {Communications Physics},
  year      = {2019},
  volume    = {2},
  number    = {1},
  month     = dec,
  doi       = {10.1038/s42005-019-0249-y},
  publisher = {Springer Science and Business Media {LLC}},
  url       = {https://doi.org/10.1038/s42005-019-0249-y},
}

\end{document}